\documentstyle[preprint,aps]{revtex}
\tighten
\begin{document}
\preprint{UCONN-97-02}
\draft
\date{\today}
\title{Gauss's law, gauge invariance, and long-range forces in QCD}
\author{Mario Belloni\thanks{e-mail address: mario@main.phys.uconn.edu},
Lusheng Chen\thanks{ e-mail address: chen@main.phys.uconn.edu}, and
Kurt Haller\thanks{ e-mail address: KHALLER@UCONNVM.UCONN.EDU}}
\address{Department of Physics, University of Connecticut, Storrs,
Connecticut 06269}
\maketitle
\begin{abstract}
We use a unitary operator constructed in earlier work to transform the
Hamiltonian for QCD in the temporal ($A_0=0$) gauge into a representation
in which the quark field is gauge-invariant, and its elementary
excitations --- quark and antiquark creation and annihilation
operators --- implement Gauss's law. In that representation,
the interactions between gauge-dependent parts of the gauge field
and the spinor (quark) field have been transformed away and replaced by
long-range non-local interactions of quark color charge densities.
These long-range interactions connect SU(3) color charge densities
through an infinite chain of gauge-invariant gauge fields either
to other SU(3) color charge densities, or to a gluon ``anchor''.
We discuss possible implications of this formalism for low-energy
processes, including confinement of quarks that are not in
color singlet configurations.
\end{abstract}
\bigskip
\bigskip

\newpage


In earlier work\cite{BCH1,CBH2}, we showed how to construct states
that implement the non-Abelian Gauss's law in QCD and Yang-Mills
theory, and we constructed gauge-invariant spinor
(quark) and gauge (gluon) operator-valued fields.  In the work
presented here, we explore the implications of these formal
developments for QCD dynamics.  In Ref.\cite{CBH2} we discussed
two unitarily equivalent representations of QCD. In
one --- the ${\cal C}$ representation --- the Gauss's
law operator in the temporal ($A_0=0$) gauge has the form
\begin{equation}
{\hat {\cal G}}^{a}({\bf r})=\partial_{i}
\Pi_{i}^{a}({\bf r})+gf^{abc}\,A_{i}^{b}({\bf r})\,
\Pi_{i}^{c}({\bf r})+j_{0}^{a}({\bf r})\;,
\label{eq:Ghat}
\end{equation}
where $\Pi_{i}^{a}({\bf r})$ is the  momentum canonically conjugate to
$A_i^{a}({\bf r})$ as well as the chromoelectric field;
$j_{0}^{a}({\bf r})$ is the quark color charge density
$j^a_{0}({\bf{r}})=g\,\,\psi^\dagger({\bf{r}})\,
(\lambda^a/2)\,\psi({\bf{r}})$, and the ${\lambda^a}$ represent
the Gell-Mann SU(3) matrices. In the unitarily equivalent ${\cal N}$
representation, that same Gauss's law operator takes the form
\begin{equation}
{\cal G}^{a}({\bf r})=D_i^{a}\Pi_{i}^{a}({\bf r})=\partial_{i}
\Pi_{i}^{a}({\bf r})+gf^{abc}\,A_{i}^{b}({\bf r})\,
\Pi_{i}^{c}({\bf r})\,.
\label{eq:G}
\end{equation}
\bigskip

The unitary equivalence that relates ${\hat {\cal G}}^{a}({\bf r})$
and ${\cal G}^{a}({\bf r})$, and that accounts for the fact that
$j_{0}^{a}({\bf r})$ does not appear explicitly
in the non-Abelian Gauss's law operator in the ${\cal N}$
representation, is similar to a relationship between two forms of the
Gauss's law operator in QED\cite{khqedtemp,khelqed}. In QED,
${\hat {\cal G}}({\bf r})=\partial_{i}\Pi_{i}({\bf r})+j_{0}({\bf r})$,
where $j_{0}({\bf r})=e\psi^{\dagger}({\bf r})\psi ({\bf r})$, and
${\hat {\cal G}}({\bf r})$ is unitarily equivalent to
$\partial_{i}\Pi_{i}({\bf r})$, which
represents the complete Gauss's law operator in what corresponds
to the ${\cal N}$ representation for QED. When all operators and
states for QED in the temporal gauge are
subjected to the unitary transformation that transforms
${\hat {\cal G}}({\bf r})$ into $\partial_{i}\Pi_{i}({\bf r})$,
the unitary operator $U=\exp[\,-{i{\int}\,\partial_{i}\Pi_{i}({\bf r})}\,
{\omega}({\bf r})\,]\,d{\bf r}$ implements the gauge transformations
$U\,A_i({\bf r})\,U^{-1}=A_i({\bf r})+\partial_i{\omega}({\bf r})$,
where ${\omega}({\bf r})$ commutes with all
operator-valued fields; and, because $\partial_{i}\Pi_{i}({\bf r})$
is the {\em complete} Gauss's law operator  in the ${\cal N}$
representation, and ${\psi}({\bf r})$ commutes with
$\partial_{i}\Pi_{i}({\bf r})$, ${\psi}({\bf r})$ represents a
gauge-invariant charged particle (electron)
field in the ${\cal N}$ representation.
\bigskip

When the Hamiltonian for QED in the temporal gauge is transformed
into the ${\cal N}$ representation, the only remaining dynamical
interactions between the electron field and the electromagnetic
field are the ones between the electrons and the gauge-invariant
excitations of the gauge field --- which, in QED, correspond to
transversely polarized photons\cite{khqedtemp,khelqed}.
The interactions which, in the original ${\cal C}$ representation,
were mediated by the longitudinal, gauge-dependent parts of the
gauge field --- the interactions mediated by the exchange of
longitudinal photon ``ghosts'' --- appear in the ${\cal N}$
representation as the non-local Coulomb interaction given by
\begin{equation}
H_c=\frac{1}{2}\int\frac{j_{0}({\bf r})\,j_{0}({\bf r^\prime})}
{4\pi|{\bf r}-{\bf r^\prime}|}\,d{\bf r}\,d{\bf r^\prime}\,.
\label{eq:Coulomb}
\end{equation}
\bigskip

The purpose of this present work is to extend this approach to
the non-Abelian QCD. There are, of course, important differences
between QED and QCD.  As was shown in Refs.\cite{khqedtemp} and
\cite{khelqed}, the states that implement Gauss's Law in the
${\cal N}$ representation for QED, $i.\,e.$ the
states that satisfy  $\partial_{i}\Pi_{i}({\bf r})
|n\rangle=0$~ --- or, more properly for QED,
$\partial_{i}\Pi_{i}^{(+)}({\bf r})|n\rangle=0$, where
$\partial_{i}\Pi_{i}^{(+)}$ selects the annihilation operator part
of $\partial_{i}\Pi_{i}$ --- constitute a Fock space that can easily be
constructed. The ${\cal N}$-representation states that satisfy
${\cal G}^{a}({\bf r})|{\nu}\rangle=0$ in QCD,
however, can only be represented as complicated coherent superpositions
of perturbative states that are neither readily orthogonalizable
nor normalizable\cite{CBH2}. Nevertheless, there are also significant
similarities between QCD and QED. In particular, since
${\hat {\cal G}}^{a}({\bf r})$ and ${\cal G}^{a}({\bf r})$ are
unitarily equivalent, and since the quark field ${\psi}({\bf r})$
commutes with $A_i^a({\bf{r}})$ and $\Pi_{i}^a({\bf r})$ and therefore
with ${\cal G}^{a}({\bf r})$, ${\psi}({\bf r})$ represents the
gauge-invariant quark field in the ${\cal N}$ representation of QCD.
And, it is possible to use the unitary equivalence demonstrated
in Ref.\cite{CBH2},
\begin{equation}
{\tilde{\xi}}^a({\bf{r}})={\cal{U}}^{-1}_{\cal{C}}\,
{\xi}^a({\bf{r}})\,{\cal{U}}_{\cal{C}}\;,
\label{eq:GgqGg}
\end{equation}
where $\xi^a$ represents some functional of operator-valued quark
and/or gluon fields in the ${\cal C}$ representation and
${\tilde{\xi}^a}$ its transform in the ${\cal N}$
representation, to transform dynamical variables from one
representation to the other. In Ref.\cite{CBH2}\,, we showed that
${\cal G}^{a}({\bf r})$ is the ${\cal N}$ representation form of
the Gauss's law operator which, in the ${\cal C}$ representation,
is given by ${\hat{\cal G}}^{a}({\bf r})\,$. Here,  we will apply this
transformation to transform the QCD Hamiltonian from the ${\cal C}$
to the ${\cal N}$ representation. In the ${\cal N}$ representation,
the excitations of the ``bare'' quark field ${\psi}({\bf r})$
are gauge-invariant, and therefore already include the gauge-dependent
gauge field components required to obey Gauss's law.  The interactions
between these components of the gauge fields and quarks therefore will
no longer appear explicitly in the Hamiltonian in the ${\cal N}$
representation, but will be replaced by non-local interactions among
quarks, which form a kind of non-Abelian analog to the Coulomb interaction
in QED.
\bigskip

We observe that
${\cal{U}}_{\cal{C}}$ is given by ${\cal{U}}_{\cal{C}}=e^{{\cal C}_{0}}
e^{\bar {\cal C}}$, where
${{\cal C}_{0}}$ and ${\bar {\cal C}}$ are
\begin{equation}
{\cal C}_{0}=i\,\int d{\bf{r}}\,
{\textstyle {\cal X}^{\alpha}}({\bf r})\,j_{0}^{\alpha}({\bf r})
\;\;\;\;\;\mbox{and}\;\;\;\;\;\;\;
{\bar {\cal C}}=i\,\int d{\bf{r}}\,
\overline{{\cal Y}^{\alpha}}({\bf r})\,j_{0}^{\alpha}({\bf r})
\label{eq:CCbar}
\end{equation}
respectively, where ${\cal{X}}^\alpha({\bf{r}}) =
[\,{\textstyle\frac{\partial_i}{\partial^2}}A_i^\alpha({\bf{r}})\,]$ and
$\overline{{\cal Y}^{\alpha}}({\bf r})=
\frac{\partial_{j}}{\partial^{2}}
\overline{{\cal A}_{j}^{\alpha}}({\bf r})$,
and where $\overline{{\cal A}_{j}^{\alpha}}({\bf r})$ is defined in
Ref.\cite{CBH2} by an operator differential equation.
We will not repeat that definition here since, for our
purposes in this work, the form of
$\overline{{\cal A}_{j}^{\alpha}}({\bf r})$ is less important than the
fact that the unitary equivalence expressed in Eq.~(\ref{eq:GgqGg}) obtains.
\bigskip

The form of the temporal gauge QCD Hamiltonian in the
${\cal C}$ representation is
\begin{eqnarray}
H &=&\int d{\bf r} \left\{\ {\textstyle \frac{1}{2}
\Pi^{a}_{i}({\bf r})\Pi^{a}_{i}({\bf r})
+ \frac{1}{4}} F_{ij}^{a}({\bf r}) F_{ij}^{a}({\bf r})\right.
\nonumber\\
\;\;\;\;\;\;\;& & +\; \left.{\psi^\dagger}({\bf r})
\left[\,\beta m-i\alpha_{i}
\left(\,\partial_{i}-igA_{i}^{a}({\bf r})
{\textstyle\frac{\lambda^\alpha}{2}}\,\right)\,\right]
\psi({\bf r})\right\}\,
\label{eq:HQCDC}
\end{eqnarray}
where
$F_{ij}^{a}({\bf r})= \partial_{j}A_{i}^{a}({\bf r})
-\partial_{i}A_{j}^{a}
({\bf r})-gf^{abc}A_{i}^{b}({\bf r})A_{j}^{c}({\bf r})$,
and where the term ${\int}\,A_0^a({\bf r})\,
{\cal G}^{a}({\bf r})\,d{\bf r}$, which appears in the
Hamiltonian when primary constraints are avoided in the
canonical quantization of this model\cite{khymtemp}, has
been omitted because it is
irrelevant for this work. It is obvious that, in transforming $H$
to the ${\cal N}$ representation, $F_{ij}^{a}({\bf r})$ remains
untransformed.  It is much more difficult, however, to transform
$\Pi^{a}_{i}({\bf r})$ to its ${\cal N}$ representation transform,
${\tilde\Pi}^{a}_{i}({\bf r})$. We have found that
${\tilde\Pi}^{a}_{i}({\bf r})$ is given by
\begin{equation}
\tilde{\Pi}^{a}_{i}({\bf r})=\Pi^{a}_{i}({\bf r})+
\sum_{m=0}\sum_{n=0}\sum_{r=0}
{\textstyle\frac{g^{m+n+r}}{m!n!}}(-1)^{m+n+r}
f^{\vec{\mu}ac}_{(m)}f^{\vec{\nu}cd}_{(n)}
f^{\vec{\delta}d h}_{(r)}
{\cal R}_{(m)}^{\vec{\mu}}({\bf r})\,
{\cal M}_{(n)}^{\vec{\nu}}({\bf r})
{\textstyle\frac{\partial_{i}}{\partial^{2}}}
\left({\cal T}_{(r)}^{\vec{\delta}}({\bf r})
j_0^{h}({\bf r})\right)\,,
\label{eq:pitilde}
\end{equation}
where $f^{\vec{\alpha}\beta\gamma}_{(\eta)}$ is
the chain of SU(3) structure functions
\begin{equation}
f^{\vec{\alpha}\beta\gamma}_{(\eta)}=f^{\alpha[1]\beta b[1]}\,
\,f^{b[1]\alpha[2]b[2]}\,f^{b[2]\alpha[3]b[3]}\,\cdots\,
\,f^{b[\eta-2]\alpha[\eta-1]b[\eta-1]}
f^{b[\eta-1]\alpha[\eta]\gamma}\;,
\label{eq:fproductN}
\end{equation}
${\cal{R}}^{\vec{\alpha}}_{(\eta)}({\bf{r}})=
\prod_{m=1}^\eta{\cal{X}}^{\alpha[m]}({\bf{r}})\,,$ and
${\cal{M}}_{(\eta)}^{\vec{\alpha}}({\bf{r}})
=\prod_{m=1}^\eta
\overline{{\cal Y}^{\alpha[m]}}({\bf{r}})\,$.
${\cal T}_{(r)}^{\vec{\delta}}({\bf r})j_0^{a}({\bf{r}})$ is defined by
\begin{equation}
{\cal T}_{(r)}^{\vec{\delta}}({\bf r})j_0^{a}({\bf{r}})=
A_{{\sf GI}\,j(1)}^{{\delta}(1)}({\bf r})\,
{\textstyle\frac{\partial_{j(1)}}{\partial^{2}}}
\left(A_{{\sf GI}\,j(2)}^{{\delta}(2)}({\bf r})\,
{\textstyle\frac{\partial_{j(2)}}{\partial^{2}}}
\left(\cdots\left(A_{{\sf GI}\,j(r)}^{{\delta}(r)}({\bf r})\,
{\textstyle\frac{\partial_{j(r)}}{\partial^{2}}}
\left(j_0^{a}({\bf{r}})\right) \right)\right)\right),
\label{eq:calT}
\end{equation}
 where $A_{{\sf GI}\,j}^{{\delta}}({\bf r})$ represents
the gauge-invariant gauge field, which can be expressed as
\begin{equation}
[\,A_{{\sf GI}\,i}^{b}({\bf{r}})\,{\textstyle\frac{\lambda^b}{2}}\,]
=V_{\cal{C}}({\bf{r}})\,[\,A_{i}^b({\bf{r}})\,
{\textstyle\frac{\lambda^b}{2}}\,]\,
V_{\cal{C}}^{-1}({\bf{r}})
+{\textstyle\frac{i}{g}}\,V_{\cal{C}}({\bf{r}})\,
\partial_{i}V_{\cal{C}}^{-1}({\bf{r}})\,;
\label{eq:AdressedAxz}
\end{equation}
$V_{\cal{C}}({\bf{r}})$ is a unitary operator given by
$V_{\cal{C}}({\bf{r}})=\exp\left(\,
-ig{\overline{{\cal{Y}}^\alpha}}({\bf{r}})
{\textstyle\frac{\lambda^\alpha}{2}}\,\right)\,
\exp\left(-ig{\cal X}^\alpha({\bf{r}})
{\textstyle\frac{\lambda^\alpha}{2}}\right)$.
We have been able to establish Eq.~(\ref{eq:pitilde}) through
an inductive proof that proceeds on very similar lines as the proof
of the ``fundamental theorem'' in Ref.\cite{CBH2}. We have further
verified Eq.~(\ref{eq:pitilde}) by showing that it is consistent with
\begin{equation}
\partial_{i}{\tilde\Pi}_{i}^{a}({\bf r})+gf^{abc}\,A_{i}^{b}({\bf r})\,
{\tilde\Pi}_{i}^{c}({\bf r})+{\tilde j}_{0}^{a}({\bf r})=\partial_{i}
\Pi_{i}^{a}({\bf r})+gf^{abc}\,A_{i}^{b}({\bf r})\,
\Pi_{i}^{c}({\bf r})\,.
\label{eq:Uconsist}
\end{equation}
Equation~(\ref{eq:pitilde}) is a non-Abelian analog to a relation that,
in QED, has the form \cite{khqedtemp,khelqed}
\begin{equation}
\tilde{\Pi}_{i}({\bf r})=\Pi_{i}({\bf r})
+\frac{\partial_{i}}{\partial^{2}}j_0({\bf r})\,,
\label{eq:piqed}
\end{equation}
In Eq.~(\ref{eq:piqed}), $j_0({\bf r})$ is the entire electric
charge density --- a consequence of the fact that, in QED,
charge resides only in the fields coupled to the gauge field,
and never in the gauge field itself.
In Eq.~(\ref{eq:pitilde}), $j_0^{a}({\bf{r}})$
represents contributions to the SU(3) color charge density
from quarks only.  The gluon color charge density, $J^a_0({\bf r})$,
does not appear explicitly in ${\tilde\Pi}_{i}^{a}({\bf r})$
(or, later in this paper, in $\tilde{H}$), because the unitary
transformation we have used to relate ${\tilde\Pi}_{i}^{a}({\bf r})$
and $\Pi_{i}^{a}({\bf r})$ does not eliminate the coupling of the
gauge-dependent parts of the gauge field to $J^a_0({\bf r})$.
\bigskip

When we transform the quark field contribution to the Hamiltonian
in the ${\cal N}$ representation, we obtain
\begin{eqnarray}
{\tilde H}_{quark}=&&{\tilde{\psi}^\dagger}({\bf r})
\left[\,\beta m-i\alpha_{i}
\left(\,\partial_{i}-igA_{i}^{a}({\bf r})\,
{\textstyle\frac{\lambda^\alpha}{2}}\,\right)\,\right]
{\tilde\psi}({\bf r})={\psi^\dagger}({\bf r})
\left[\,\beta m-i\alpha_{i}\partial_{i}\,\right]
\psi({\bf r})
\nonumber \\
-i&&{\psi^\dagger}({\bf r})\alpha_{i}\left[\,V_{\cal{C}}({\bf{r}})\,
\partial_{i}V_{\cal{C}}^{-1}({\bf{r}})
-igV_{\cal{C}}({\bf{r}})\left(\,A_{i}^b({\bf{r}})\,
{\textstyle\frac{\lambda^b}{2}}\,\right)
V_{\cal{C}}^{-1}({\bf{r}})\,\right]\psi({\bf r})\,
\label{eq:psiham};
\end{eqnarray}
and, from
\begin{equation}
\tilde{\psi}({\bf r}) = V_{\cal{C}}^{-1}({\bf{r}})\,\psi({\bf r})
\;\;\;\;\;\;\;\;\;\;\;\;\mbox{and}
\;\;\;\;\;\;\;\;\;\;\;\;\tilde{\psi}^\dagger({\bf{r}}) =
\psi^\dagger({\bf{r}})\,V_{\cal{C}}({\bf{r}})\,,
\label{eq:psivc}
\end{equation}
and from Eqs.~(\ref{eq:AdressedAxz}) and (\ref{eq:psiham}),
\begin{equation}
 {\tilde H}_{quark}={\psi^\dagger}({\bf r})\left[\,\beta m-i\alpha_{i}
\left(\,\partial_{i}-igA_{{\sf GI}\,i}^{a}({\bf r})
{\textstyle\frac{\lambda^\alpha}{2}}\,\right)\,\right]\psi({\bf r})\,.
\label{eq:psihamtwo}
\end{equation}
After the transformations from the ${\cal C}$ representation to the
${\cal N}$ representation have been carried out on all constituent
parts of $H$, we obtain
\begin{eqnarray}
\tilde{H}&=&\int d{\bf r}\left\{ \ {\textstyle \frac{1}{2}}
\Pi^{a}_{i}({\bf r})\Pi^{a}_{i}({\bf r})
+  {\textstyle \frac{1}{4}} F_{ij}^{a}({\bf r}) F_{ij}^{a}({\bf r})+
{\psi^\dagger}({\bf r})\left[\beta m-i\alpha_{i}
\left(\partial_{i}-igA_{{\sf GI}\,i}^{a}({\bf r})
{\textstyle\frac{\lambda^\alpha}{2}}\right)\right]
\psi({\bf r})\right.
\nonumber\\
&&\;\;\;\;\;\;+ {\textstyle \frac{1}{2}}\sum_{m=0}
\sum_{n=0}\sum_{r=0}
{\textstyle\frac{g^{m+n+r}}{m!n!}}(-1)^{m+n+r}
f^{\vec{\mu}ac}_{(m)}f^{\vec{\nu}
cd}_{(n)}f^{\vec{\delta}d h}_{(r)}\Pi^{a}_{i}({\bf r})
{\cal R}_{(m)}^{\vec{\mu}}({\bf r})\,
{\cal M}_{(n)}^{\vec{\nu}}({\bf r})
{\textstyle\frac{\partial_{i}}{\partial^{2}}}
\left({\cal T}_{(r)}^{\vec{\delta}}({\bf r})
j_0^{h}({\bf r})\right)
\nonumber \\
&&\;\;\;\;\;\;+ {\textstyle \frac{1}{2}}
\sum_{m=0}\sum_{n=0}\sum_{r=0}
{\textstyle\frac{g^{m+n+r}}{m!n!}}
(-1)^{m+n+r}f^{\vec{\mu}ac}_{(m)}
f^{\vec{\nu}cd}_{(n)}f^{\vec{\delta}d h}_{(r)}
{\cal R}_{(m)}^{\vec{\mu}}({\bf r})\,
{\cal M}_{(n)}^{\vec{\nu}}({\bf r})
{\textstyle\frac{\partial_{i}}{\partial^{2}}}
\left({\cal T}_{(r)}^{\vec{\delta}}({\bf r})
j_0^{h}({\bf r})\right)\Pi^{a}_{i}({\bf r})\;
\nonumber \\
&&\;\;\;\;\;\;\left.+ {\textstyle \frac{1}{2}}
\sum_{r=0}\sum_{r^\prime =0}
g^{r+r^\prime}(-1)^{r+r^\prime}
f^{\vec{\delta}d h}_{(r)}
f^{\vec{\delta}^\prime d h^\prime}_{(r^\prime)}
{\textstyle\frac{\partial_{i}}{\partial^{2}}}
\left({\cal T}_{(r)}^{\vec{\delta}}({\bf r})
j_0^{h}({\bf r})\right)
{\textstyle\frac{\partial_{i}}{\partial^{2}}}
\left({\cal T}_{(r^\prime)}^{\vec{\delta}^\prime}({\bf r})
j_0^{h^\prime}({\bf r})\right)\right\}\,.
\label{eq:HQCDN}
\end{eqnarray}
In deriving Eq.~(\ref{eq:HQCDN}), we have used the identity
\begin{eqnarray}
&&\sum_{m=0}{\textstyle\frac{g^{m}}{m!}}(-1)^{m}
f^{\vec{\mu}ac}_{(m)}\,{\cal R}_{(m)}^{\vec{\mu}}({\bf r})\,
\sum_{m^\prime =0}
{\textstyle\frac{g^{m^\prime }}{m^\prime !}}
(-1)^{m^\prime}f^{\vec{\mu}^\prime a c^\prime }_{(m^\prime )}\,
{\cal R}_{(m^\prime)}^{\vec{\mu}^\prime }({\bf r})
\nonumber \\
&&\;\;\;\;\;\;=\sum_{M=0}\sum_{m=0}^{M}
{\textstyle\frac{g^{M}}{m!(M-m)!}}(-1)^{m+1}
f^{\vec{\mu}c^\prime c}_{(M)}\,
{\cal R}_{(M)}^{\vec{\mu}}({\bf r})\,
\label{eq:RidentA}\,.
\end{eqnarray}
Since
${\phi}(M)=\sum_{m=0}^{M}
{\textstyle\frac{M!}{m!(M-m)!}}(-1)^{m}=(1-1)^M=0$ for
$M>0$, while ${\phi}(0)=1\,$,
Eq.~(\ref{eq:RidentA}) reduces to
\begin{equation}
\sum_{m=0}{\textstyle\frac{g^{m}}{m!}}(-1)^{m}f^{\vec{\mu}ac}_{(m)}\,
{\cal R}_{(m)}^{\vec{\mu}}({\bf r})\,
\sum_{m^\prime =0}{\textstyle\frac{g^{m^\prime }}
{m^\prime !}}(-1)^{m^\prime}
f^{\vec{\mu}^\prime a c^\prime }_{(m^\prime )}\,
{\cal R}_{(m^\prime)}^{\vec{\mu}^\prime }({\bf r})
=\delta_{c,c^\prime}
\label{eq:RidentB}\,.
\end{equation}
Equation~(\ref{eq:RidentB}) also obtains when
${\cal M}_{(m)}^{\vec{\mu}}({\bf r})$
is substituted for ${\cal R}_{(m)}^{\vec{\mu}}({\bf r})$.
\bigskip

We can use the Baker-Hausdorff-Campbell (BHC) theorem to
show that
\begin{equation}
V_{\cal{C}}^{-1}({\bf{r}})\,{\textstyle\frac{\lambda^d}{2}}\,
V_{\cal{C}}({\bf{r}})=\sum_{m=0}\sum_{n=0}
{\textstyle\frac{g^{m+n}}{m!n!}}(-1)^{m+n}
f^{\vec{\mu}ac}_{(m)}f^{\vec{\nu} cd}_{(n)}\,
{\cal R}_{(m)}^{\vec{\mu}}({\bf r})\,
{\cal M}_{(n)}^{\vec{\nu}}({\bf r})\,
{\textstyle\frac{\lambda^a}{2}}\;,
\label{eq:RMVC}
\end{equation}
and, from Eqs.~(\ref{eq:AdressedAxz}) and (\ref{eq:RMVC}),
\begin{eqnarray}
&&\sum_{m=0}\sum_{n=0}
{\textstyle\frac{g^{m+n}}{n!m!}}
(-1)^{m+n}f^{\vec{\mu}ac}_{(m)}
f^{\vec{\nu}cd}_{(n)}\,
\partial_i\left(\,{\cal{R}}_{(m)}^{\vec{\mu}}\,
{\cal{M}}_{(n)}^{\vec{\nu}}\,\right)V_i^d
\nonumber\\
&&=g\sum_{m=0}\sum_{n=0}
{\textstyle\frac{g^{m+n}}{n!m!}}
(-1)^{m+n}f^{\vec{\mu}ab}_{(m)}
f^{\vec{\nu}bc}_{(n)}
f^{c\sigma e}\,{\cal{R}}_{(m)}^{\vec{\mu}}\,
{\cal{M}}_{(n)}^{\vec{\nu}}\,A_{{\sf GI}\,i}^{\sigma}({\bf{r}})\,
V_i^e
\nonumber\\
&&-gA_i^\tau\,f^{a\tau b}\sum_{m=0}\sum_{n=0}
{\textstyle\frac{g^{m+n}}{n!m!}}(-1)^{m+n}f^{\vec{\mu}bc}_{(m)}
f^{\vec{\nu}cd}_{(n)}\,{\cal{R}}_{(m)}^{\vec{\mu}}\,
{\cal{M}}_{(n)}^{\vec{\nu}}\,V_i^d
\label{eq:LCconj}\,.
\end{eqnarray}
And we can use Eqs.~(\ref{eq:RMVC}) and (\ref{eq:LCconj}) to
reexpress Eq.~(\ref{eq:HQCDN}) in the form
\begin{eqnarray}
\tilde{H}&=&\int d{\bf r}\left\{\,{\textstyle \frac{1}{2}}
\Pi^{a}_{i}({\bf r})\,\Pi^{a}_{i}({\bf r})
+  {\textstyle \frac{1}{4}} F_{ij}^{a}({\bf r})
F_{ij}^{a}({\bf r})+{\psi^\dagger}({\bf r})
\left[\,\beta m-i\alpha_{i}\left(\,\partial_{i}
-igA_{{\sf GI}\,i}^{a}({\bf r})
{\textstyle\frac{\lambda^\alpha}{2}}\,\right)\,\right]
\psi({\bf r})\right\}
\nonumber\\
&&\;\;\;\;\;\;+{\tilde{H}}_{\cal G}+{\tilde{H}}_{LR}
\label{eq:HQCDNVC}
\end{eqnarray}
with
\begin{eqnarray}
{\tilde{H}}_{\cal G}&=&\int d{\bf r}\left\{\, -{\sf Tr}
\left[\,\sum_{r=0}g^{r}(-1)^{r}f^{\vec{\delta}d h}_{(r)}
{\cal G}^{a}({\bf r}){\textstyle\frac{\lambda^a}{2}}
V_{\cal{C}}^{-1}({\bf{r}})
{\textstyle\frac{\lambda^d}{2}}V_{\cal{C}}({\bf{r}})
{\textstyle\frac{\partial_{i}}{\partial^{2}}}
\left({\cal T}_{(r)}^{\vec{\delta}}({\bf r})
j_0^{h}({\bf r})\right)\,\right]\right.\;
\nonumber \\
&&\;\;\;\;\;\;\left.- {\sf Tr}\left[\,\sum_{r=0}
g^{r}(-1)^{r}f^{\vec{\delta}d h}_{(r)}
{\textstyle\frac{\partial_{i}}{\partial^{2}}}
\left({\cal T}_{(r)}^{\vec{\delta}}({\bf r})
j_0^{h}({\bf r})\right)V_{\cal{C}}^{-1}({\bf{r}})
{\textstyle\frac{\lambda^d}{2}}V_{\cal{C}}({\bf{r}})
{\cal G}^{b}({\bf r})
{\textstyle\frac{\lambda^b}{2}}\,\right]\,\right\}\;
\label{eq:HQCDNVCG}
\end{eqnarray}
and
\begin{eqnarray}
{\tilde{H}}_{LR}&=&\int d{\bf r}\left\{\,+{\sf Tr}\left[\,\sum_{r=0}
g^{r+1}(-1)^{r}f^{\vec{\delta}d h}_{(r)}f^{d\sigma e}
\Pi^{a}_{i}({\bf r})\,{\textstyle\frac{\lambda^a}{2}}\,
V_{\cal{C}}^{-1}({\bf{r}})
{\textstyle\frac{\lambda^e}{2}}V_{\cal{C}}({\bf{r}})
A_{{\sf GI}\,i}^{\sigma}({\bf r})
{\textstyle\frac{\partial_{i}}{\partial^{2}}}
\left({\cal T}_{(r)}^{\vec{\delta}}({\bf r})
j_0^{h}({\bf r})\right)\,\right]\right.\;
\nonumber \\
&&\;\;\;\;\;\;+ {\sf Tr}\left[\,\sum_{r=0}
g^{r+1}(-1)^{r}f^{\vec{\delta}d h}_{(r)}f^{d\sigma e}
{\textstyle\frac{\partial_{i}}{\partial^{2}}}
\left({\cal T}_{(r)}^{\vec{\delta}}({\bf r})
j_0^{h}({\bf r})\right)V_{\cal{C}}^{-1}({\bf{r}})
{\textstyle\frac{\lambda^e}{2}}V_{\cal{C}}({\bf{r}})
A_{{\sf GI}\,i}^{\sigma}({\bf r})
\Pi^{b}_{i}({\bf r}){\textstyle\frac{\lambda^b}{2}}\,\right]\;
\nonumber \\
&&\;\;\;\;\;\;+ \left.{\textstyle \frac{1}{2}}\sum_{r=0}
\sum_{r^\prime =0}
g^{r+r^\prime}(-1)^{r+r^\prime}
f^{\vec{\delta}d h}_{(r)}f^{\vec{\delta}^\prime d h^\prime}_{(r^\prime)}
{\textstyle\frac{\partial_{i}}{\partial^{2}}}
\left({\cal T}_{(r)}^{\vec{\delta}}({\bf r})
j_0^{h}({\bf r})\right)
{\textstyle\frac{\partial_{i}}{\partial^{2}}}
\left({\cal T}_{(r^\prime)}^{\vec{\delta}^\prime}({\bf r})
j_0^{h^\prime}({\bf r})\right)\,\right\}\;.
\label{eq:HQCDNVCRL}
\end{eqnarray}
\bigskip

It is instructive to examine the physical significance of the
individual terms in Eq.~(\ref{eq:HQCDNVC}).  We observe that
$\tilde{H}$ includes the kinetic energy of the quark and gluon
fields, as well as an interaction term between the spatial
SU(3) current term for the  spinor field (which is gauge-invariant
in the ${\cal N}$ representation) and the gauge-invariant
gauge field $A_{{\sf GI}\,i}^{a}({\bf r})$, which is gauge-invariant
in {\em both} the ${\cal C}$ and the ${\cal N}$ representations.
$\tilde{H}$ also includes $\tilde{H}_{\cal G}$, in which the
Gauss's law operator ${\cal G}$ appears on either
the extreme left-hand or right-hand side of an operator product.
$\tilde{H}_{\cal G}$ vanishes when evaluated between
states that implement Gauss's law in the ${\cal N}$ representation.
$\tilde{H}$ also includes a long-range non-local interaction,
$\tilde{H}_{LR}$, which consists of three terms; the last of
these three terms in Eq.~(\ref{eq:HQCDNVCRL}) describes the interaction
between two SU(3) color charge densities through a non-local interaction
that can be characterized as a non-Abelian analog of the Coulomb interaction
in QED; but it differs from the Coulomb interaction in QED in that the factor
$(1/2)\,{\partial}^{-2}$ does not connect two $j_0^{a}$ terms directly. The
long-range interaction in this case is transmitted between the two quark
color charge densities through an infinite chain of gauge-invariant gauge
fields as shown in Eq.~(\ref{eq:calT}). The other two terms in
Eq.~(\ref{eq:HQCDNVCRL})  each connect a $j_0^{a}$ term through an identical
chain of gauge-invariant gauge fields to the gluon field itself,
``anchoring'' the $j_0^{a}$ term, finally, in the gluon field through an
$A_{{\sf GI}\,i}^{a}({\bf r})\,\Pi^{b}_{i}({\bf r})$ term.
\bigskip

It is appealing to speculate that the three terms in Eq.~(\ref{eq:HQCDNVCRL})
have an important role in confining quark states that are not in singlet color
configurations.  To be sure, such an idea is speculative.  The spatial
dependence of the long-range interactions in $\tilde{H}_{LR}$ cannot be
determined without more information about matrix elements of gauge fields than
we have presented here. Partial insight into these long-range interactions
can be obtained by noting that in the virtually  identical SU(2) case,
the summation over all orders, $s$, of
${\cal T}_{(s)}^{\vec{\delta}}({\bf r})$ can be performed exactly to give
\begin{equation}
\sum_{r=0}
g^{r}(-1)^{r}\epsilon^{\vec{\delta}d h}_{(r)}\,
{\cal T}_{(r)}^{\vec{\delta}}({\bf r})=-\delta_{d,h}
-{\textstyle\frac{g}{1+{\cal W}}}\epsilon^{\delta d h}\,
{\cal T}^{\delta}+{\textstyle \frac{g^2}{1+{\cal W}}}
\epsilon^{{\delta}(1) db}\epsilon^{b{\delta}(2) h}\,
{\cal T}^{{\delta}(1)}\,{\cal T}^{{\delta}(2)}\;,
\label{eq:TSU2}
\end{equation}
where ${\cal T}^{\delta}({\bf r})=
A_{{\sf GI}\,j}^{\delta}({\bf r})\,
{\textstyle\frac{\partial_{j}}{\partial^{2}}}$
and ${\cal W}=g^2({\cal T}^{\beta}{\cal T}^{\beta})\,$.  ${\cal W}$ is a
differential operator in which the $\partial_{j}/\partial^2$ in each
${\cal T}^{\beta}$ operates on {\em all} operators to its right in the
series expansion of $1/(1+{\cal W})$, as shown in Eq.~(\ref{eq:calT}).
Equation~(\ref{eq:TSU2}) therefore describes a differential operator rather
than an algebraic quantity. Nevertheless, it is a reasonable conjecture
that $ W^{(n)}$ might develop expectation values in the gluon ``sea'' in
which quarks are embedded that have the effect of turning $1/(1+{\cal W})$ into
a confining
potential, and that a very similar phenomenon might occur in the SU(3) case.
We consider this scenario to be an interesting conjecture that warrants
further study and development.
\bigskip

It is of particular interest to consider Eq.~(\ref{eq:HQCDNVCRL}) in
conjunction with a model  suitable for low-energy QCD, that combines
gluons and ``static'' quarks --- $i.\,e.$ a model in which quark kinetic
energy is small, and quark-antiquark creation and annihilation is
quenched --- but in which Gauss's law is obeyed. This model is of
interest because it allows us to exploit a pertinent analogy between
QCD and QED.  In QED, the covariant gauge formulation enjoys a clear
advantage for perturbative calculations --- particularly when radiative
corrections are required --- but is not as well adapted for extracting the
Coulomb interaction as the dominant one for low-energy phenomena, such as the
energy levels of the bound states of the Hydrogen atom.  In contrast,
in a formulation  of QED in which Gauss's law has been implemented and
the ``pure gauge'' components of the gauge field have been transformed
away, the dominance of the Coulomb interaction for low-energy phenomena
becomes very explicit\cite{{khqedtemp},{khelqed}}. Similarly, in QCD,
when we eliminate the interaction between the gauge-dependent parts of
the gauge fields and the quarks, the long-range interactions of quark
color charge densities described in Eq.~(\ref{eq:HQCDNVCRL}) become
very explicit. As in the case of QED, these long-range interactions are
not manifest in either the ${\cal C}$-representation Hamiltonian described
in Eq.~(\ref{eq:HQCDC}), or in the covariant-gauge formulation of QCD.
It is very natural to suspect that these long-range interactions
are important for low-energy processes, and that they have a connection
with the color confinement observed in low-energy phenomena. A number of
authors have suggested that there is some connection between gauge
invariance and color confinement in non-Abelian gauge
theories\cite{johnson,johnson2,johnson3,johnson4,stoll,lenz1,lenz2}.
The following observations about the application of ${\tilde{H}}_{LR}$,
the long-range part of the transformed Hamiltonian ${\tilde{H}}$, are
pertinent to this question.
\bigskip

We consider the expectation value
$\langle{\sf A}|\,j_0^{a}({\bf r})\,|{\sf A}\rangle\,$,
where $|{\sf A}\rangle\,$ is an $n$-particle state of ``static'' quarks,
in the sense defined above. We also assume that the quarks and antiquarks
are described by a set of wave functions $u^{\alpha}_{n}({\bf r})$ and
$v^{\alpha}_{n}({\bf r})$ respectively, and that the quark field
$\psi({\bf r})$ is represented as
\begin{equation}
\psi({\bf r})  =  \sum_{\bf k}\,\left[\,q^{\alpha}_{n}\,
u^{\alpha}_{n}({\bf r}) +{\bar q}^{\dagger\,{\alpha}}_{n}\,
v^{\alpha}_{n}({\bf r})\,\right]\,,
\label{eq:fieldquark}
\end{equation}
where $q^{\alpha}_{n}$ and ${\bar q}^{\dagger\,{\alpha}}_{n}$
annihilate quarks and create antiquarks respectively, and
$u^{\alpha}_{n}({\bf r})$ and $v^{\alpha}_{n}({\bf r})$ are the
positive and negative energy solutions of a Dirac equation
in an external potential ($\alpha$ labels the SU(3) index and
$n$ the energy, angular momentum, etc. of quarks
and antiquarks in the corresponding orbital states).
We will identify $|{\sf A}\rangle\,$ as a multiquark state
$|q^{{\alpha}}_{n}{\cdots} q^{{\alpha^{\prime}}}_{n^{\prime}}\rangle\,$.
We observe that for such a state,
$\langle{q^{{\alpha}}_{n}{\cdots}
q^{{\alpha^{\prime}}}_{n^{\prime}}}
|\,{\int}\,d{\bf r}\,j_0^{a}({\bf r})\,|
q^{{\alpha}}_{n}{\cdots}q^{{\alpha^{\prime}}}_{n^{\prime}}\rangle=0$ when
$|q^{{\alpha}}_{n}{\cdots}q^{{\alpha^{\prime}}}_{n^{\prime}}\rangle\,$
is a color singlet state, since ${\int}\,d{\bf r}\,j_0^{a}({\bf r})$ is the
generator of infinitesimal rotations in the SU(3) space, and the singlet
state is the scalar in that space. When
${\cal T}_{(q)}^{\vec{\delta}}({\bf r})$ imposes only minimal variations
on $j_0^{h}({\bf r})$ over the region in ${\bf r}$
in which $u^{\alpha}_{n}({\bf r})$ and $v^{\alpha}_{n}({\bf r})$
are localized, then the expectation values that enter into
the evaluation of long-range forces between ``static'' quark states in
color-singlet configurations can be expected to vanish.  For example,
when two singlet states, $|{\sf A}\rangle=|q^{{\alpha}}_{n}{\cdots}
q^{{\alpha^{\prime}}}_{n^{\prime}}\rangle\,$ and $|{\sf B}\rangle
=|Q^{{\alpha}}_{n}{\cdots}Q^{{\alpha^{\prime}}}_{n^{\prime}}\rangle\,$
interact through the last term in Eq.~(\ref{eq:HQCDNVCRL}), where
the $q^{{\alpha}}_{n}$ annihilate quarks centered about point ${\bf R}$,
and the $Q^{{\alpha}}_{n}$ annihilate quarks centered about
${\bf R}^\prime$, and where $|{\bf R}-{\bf R}^{\prime}|\gg|\delta{\bf r}|$
where $|\delta{\bf r}|$ is the mean radius of the wave functions
$u^{\alpha}_{n}({\bf r})$ and $v^{\alpha}_{n}({\bf r})$, then the
following conjecture becomes  plausible: Because the distance between
the end points of the chain of gauge-invariant gauge fields in
Eq.~(\ref{eq:HQCDNVCRL})  is much larger than $|\delta{\bf r}|\,$,
$f^{\vec{\delta} a h}_{(s)}{\cal T}_{(s)}^{\vec{\delta}}({\bf r})
j_0^{h}({\bf r})$ would terminate essentially in
${\kappa}\,{\int}\,d{\bf r}\,j_0^{a}({\bf r})\,$, where $\kappa$
is some constant. In that case, the matrix element
$f^{\vec{\delta} a h}_{(s)}{\cal T}_{(s)}^{\vec{\delta}}({\bf r})
\langle{q^{{\alpha}}_{n}{\cdots}
q^{{\alpha^{\prime}}}_{n^{\prime}}}|\,j_0^{h}({\bf r})\,
|q^{{\alpha}}_{n}{\cdots}q^{{\alpha^{\prime}}}_{n^{\prime}}\rangle$ and the
corresponding chain terminating in the quark state
$|Q^{{\alpha}}_{n}{\cdots}Q^{{\alpha^{\prime}}}_{n^{\prime}}\rangle\,$ both
would vanish.  When $|{\bf R}-{\bf R}^{\prime}|$ decreases,
the cancellation can become less complete, and ``van der Waals'' forces
between singlet states become a reasonable conjecture.
\bigskip

This account of how confining long-range forces might develop
between multiquark states that are not in color singlet configurations,
is of course not a proof. Much further work is required to support such
a scenario with quantitative arguments --- for example, it is not clear
from this discussion that an attractive rather than a repulsive force
will develop, even if ${\tilde{H}}_{LR}$ does become unboundedly
large for large separations between configurations of quarks that
are not in color singlet configurations. Whether or not these scenarios
for quark color confinement are borne out by further investigations,
${\tilde{H}}\,$ --- the QCD Hamiltonian in the ${\cal N}$ representation
of the temporal gauge --- promises to be a productive formulation of
QCD dynamics for further investigations into the low-energy properties of QCD.
\bigskip

This research was supported by the Department of Energy
under Grant No. DE-FG02-92ER40716.00.

\end{document}